\documentclass[conference]{IEEEtran}

\usepackage{array,booktabs}
\usepackage{hyperref}

\usepackage{cite}

\ifCLASSINFOpdf
  \usepackage[pdftex]{graphicx}
\else
  \usepackage[dvips]{graphicx}
\fi
\usepackage{amsmath}
\begin{document}
%
\title{Designing Pre-training Datasets from Unlabeled Data for EEG Classification with Transformers}

\author{\IEEEauthorblockN{Tim Bary, Benoît Macq}
\IEEEauthorblockA{Universit\'e catholique de Louvain\\
ICTEAM, ELEN\\
Louvain-la-Neuve, Belgium\\
\{tim.bary\}\{benoit.macq\}@uclouvain.be}}



%


\maketitle

\begin{abstract}
Transformer neural networks require a large amount of labeled data to train effectively. Such data is often scarce in electroencephalography, as annotations made by medical experts are costly. This is why self-supervised training, using unlabeled data, has to be performed beforehand.
In this paper, we present a way to design several labeled datasets from unlabeled electroencephalogram (EEG) data. These can then be used to pre-train transformers to learn representations of EEG signals. We tested this method on an epileptic seizure forecasting task on the Temple University Seizure Detection Corpus using a Multi-channel Vision Transformer. Our results suggest that 1) Models pre-trained using our approach demonstrate significantly faster training times, reducing fine-tuning duration by more than 50\% for the specific task, and 2) Pre-trained models exhibit improved accuracy, with an increase from 90.93\% to 92.16\%, as well as a higher AUC, rising from 0.9648 to 0.9702 when compared to non-pre-trained models.
\end{abstract}


\IEEEpeerreviewmaketitle

\section{Introduction}
Transformer neural networks distinguish themselves from other deep learning architectures thanks to their self-attention mechanism. This feature allows these networks to focus on different parts of the input sequence as they process it, facilitating the capture of long-range dependencies and contextual relationships. On top of this, this type of network also operates on the entire sequence of input data at once, allowing for parallelization and improved efficiency. These traits however come at a price: transformer-based models require a substantial amount of annotated data during training to avoid overfitting \cite{vaswani2017,vig2019}.

While transformers have primarily found their footing in Natural Language Processing (NLP) \cite{vaswani2017} and Computer Vision (CV) \cite{dosov2020}, researchers have recently started extending their capabilities to electroencephalogram (EEG) applications, including emotion recognition \cite{wang2022}, motor imagery \cite{song2022,xie2022}, and epileptic seizure forecasting \cite{hussein2022,yan2022}. However, unlike for NLP and CV, which have a lot of labeled data available to train their models, quality annotated data in the medical field is scarce and has a high acquisition cost, since it requires an expertise exclusive to medical practitioners. 

To address this challenge, Self-Supervised Learning (SSL) emerges as a pivotal approach. SSL harnesses the available unlabeled data in a specific field and generates pseudo-labels for it through some of its intrinsic characteristics. The model is then pre-trained on this “self-labeled” data, enabling it to acquire some fundamental data representation. Subsequently, these pre-trained models can be fine-tuned for specific applications using a small sample of labeled data, outperforming their non-pre-trained counterparts \cite{zhang2000,torrey2010, spathis2022}.

In this article, we present a methodology to exploit the huge amount of unlabeled EEG data available by designing three intuitive and interpretable pre-training tasks. We benchmark these pre-trainings by comparing their contributions to the performances of Hussein \textit{et al.}'s Multi-channel Vision Transformer (MViT) \cite{hussein2022} on an eyes open/eyes closed (EO/EC) classification dataset. Finally, we test the performance improvement of the best pre-training on an epileptic seizure forecasting task, leveraging the Temple University Hospital Seizure Detection Corpus (TUSZ) \cite{shah2018}.

\section{Pre-training datasets}
We collected two datasets to demonstrate the capabilities of the proposed methods. Their specifications are described in subsection \ref{sec:dataset}. The main contribution of this article, namely the processes by which unlabeled data from these datasets is converted into useful pre-training data, is described in subsection \ref{SEC:Alt}.

\subsection{The initial datasets}\label{sec:dataset}
\begin{figure*}[!t]
    \centering
    \includegraphics[width=\textwidth]{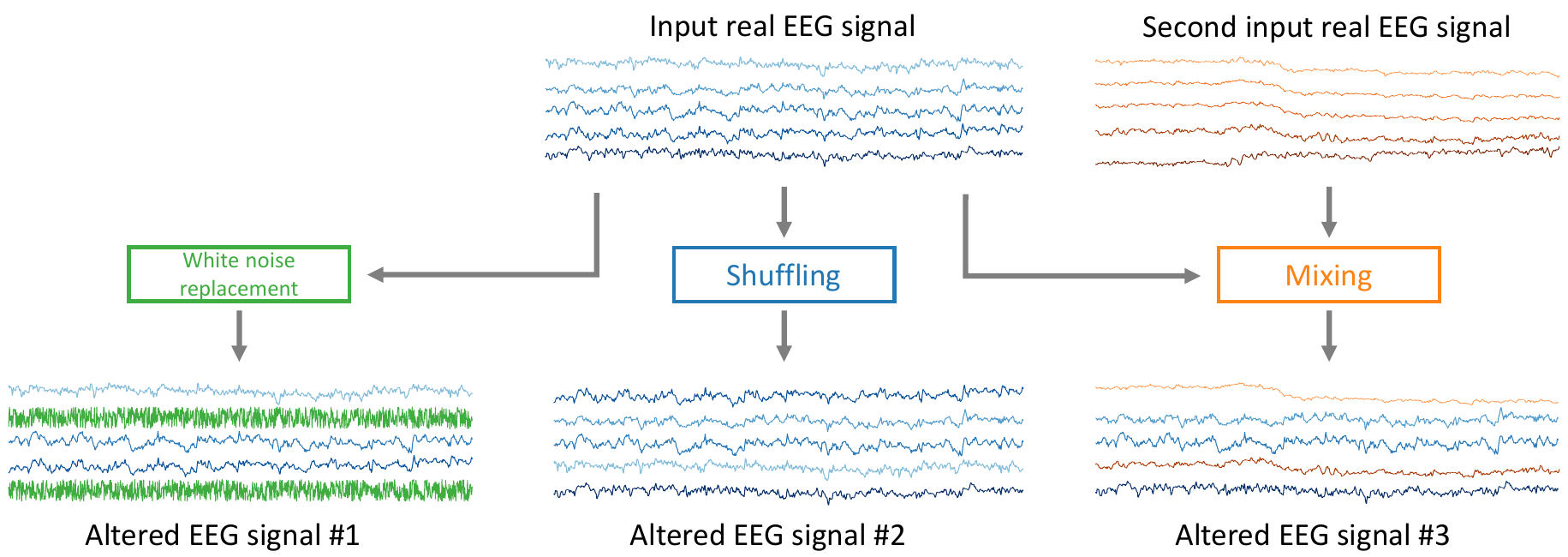}
    \caption{Visualisation of the three EEG signals alterations for the proposed pre-training tasks. Alteration \#1 replaces channels of the EEG with white noise, alteration \#2 shuffles the order of the channels, and alteration \#3 mixes two randomly paired EEG samples together by replacing the channels of the first sample by the channels of the second one and vice versa.}
    \label{FIG1}
\end{figure*}
\subsubsection{The EO/EC dataset}
This dataset features a healthy male subject in the 20-25 age category. It includes an 11 minutes, 32 channels EEG signal recorded at 1 kHz. Annotations comprise two labels: EO when the subject has his eyes open while his EEG is being recorded, EC when his eyes are closed. Given its small size and the simplicity of the task, the EO/EC dataset was used for experimenting with the different pre-training methods to identify the most effective one with statistically significant results.

\subsubsection{The TUSZ dataset}
The TUSZ database is part of a larger EEG corpus created to facilitate EEG processing tool development \cite{obeid2016}. TUSZ comprises EEG signals from 579 patients in 1175 sessions, including 352 sessions with seizures. Annotations exploited in this paper consist of the start and end times of seizures during the session, although more detailed information is available. Specifically, we leveraged these elements to generate our classification labels, that is, PI for pre-ictal, and II for inter-ictal. The first label designates EEG signals present from a few minutes to a few seconds before a seizure, while the second is used to describe patterns that do not immediately precede or follow a seizure. Being able to segregate both labels therefore allows to forecast epileptic seizures.

Four unipolar EEG montages were employed in the dataset, two using an average reference and the other two including earlobe-referenced electrodes. Half of the montages use 19 electrodes, and the rest use 21, all following the 10-20 system \cite{jasper1958}. In this paper, the TUSZ corpus is employed for benchmarking against a literature-standard task and evaluating the impact of pre-training on model performances in a practical scenario.

To simulate the scarcity of labeled data, we randomly excluded $70\%$ of the labels on both datasets. This share of unlabeled data is the one used to generate the pre-training datasets. The remaining $30\%$ of labeled data is reserved for fine-tuning on the respective specific tasks.

\subsection{Altering the EEG signals} \label{SEC:Alt}
As previously explained, the goal of a pre-training task is to draw the model to already internalize some key features of the kind of data it is being trained on. Doing so allows the initial weights of the model to be closer to the ones minimizing the loss function when training on the useful task, saving time and increasing performances \cite{torrey2010, spathis2022}. Keeping this idea in mind, we came up with three different ways to teach EEG properties to a deep learning model. 

All three proposed methods can be summarized as binary classification tasks, distinguishing between "EEG" and "non-EEG" multi-channels signals. More specifically, we collected unlabeled EEG data that we split in two halves, keeping one part as the control, real EEG, and altering the other part to become the "non-EEG" signal. Three relevant alterations have been designed so far, characterizing the three methods depicted underneath and illustrated in Fig. \ref{FIG1}.

\subsubsection{White noise replacement}
In this first type of modification, $n$ random channels of the original EEG are replaced with Gaussian white noise. The number of replaced channels $n$ is a random integer varying uniformly between $1$ and hyperparameter $N$ for each sample generated. It is expected that a higher number of channels affected leads to more ease for the model at recognizing the "non-EEG" samples. $N$ therefore controls the "difficulty" of the task (the closer $N$ is to 1, the harder the classification becomes), and a value of $5$ has been arbitrarily selected. The intuition behind this modification is that, unlike with EEG signals that have a power spectral density which tends to decrease with the frequency \cite{namazi2012}, the spectral density of white noise remains constant across all frequencies. Being able to part between white noise and EEG frequency behaviors therefore implies that the model has integrated such behaviors.
\subsubsection{Shuffling}
In this modification, all the channels are conserved, but their positions are permuted. This operation amounts to shuffling the locations of the different EEG probes for the model, and it therefore has to learn the inter-dependencies between the channels to classify the data as shuffled or not. Channels in EEG are correlated proportionally to the distance between them. That is, the further they are located from one another, the less correlated they are \cite{bhav2018}.
\subsubsection{Mixing}
This third method takes two EEG samples as input and selects $n \in [1,N]$ channels from the first input to replace channels in the second input and vice versa. The objective is to have channels that are not correlated to any other channel, although having the same behavior as normal EEG signals. Like with channel shuffling, this classification task aims at teaching the channels correlations to the model. Once again, the hyperparameter $N$ has arbitrarily been set to $5$ (in this case, the further away $N$ is from $n_{channels}/2 = $ $16$ for EO/EC and $10$ for TUSZ, the more difficult the task becomes, as less uncorrelated channels are potentially identifiable).

\section{Design of the model}
\subsection{Transformer architecture}
We selected the Multi-channel Vision Transformer (MViT) architecture from Hussein \textit{et al.} \cite{hussein2022} to benchmark our pre-trainings. This architecture, designed specifically for EEG classification, was adopted because its configuration makes it more parameter extensive than a majority of the other approaches, meaning that this model would benefit the most from our contribution compared to less bulky models. The MViT relies on Dosovitskiy \textit{et al.}'s Vision Transformer \cite{dosov2020}, which performs image classification by partitioning an image into non-overlapping patches, flattening them, and feeding them through an encoder after positional embedding. The encoder output is then processed by a Multi-Layer Perceptron (MLP) to classify the image. 

\begin{figure}[!t]
    \centering
    \includegraphics[width=3.5in]{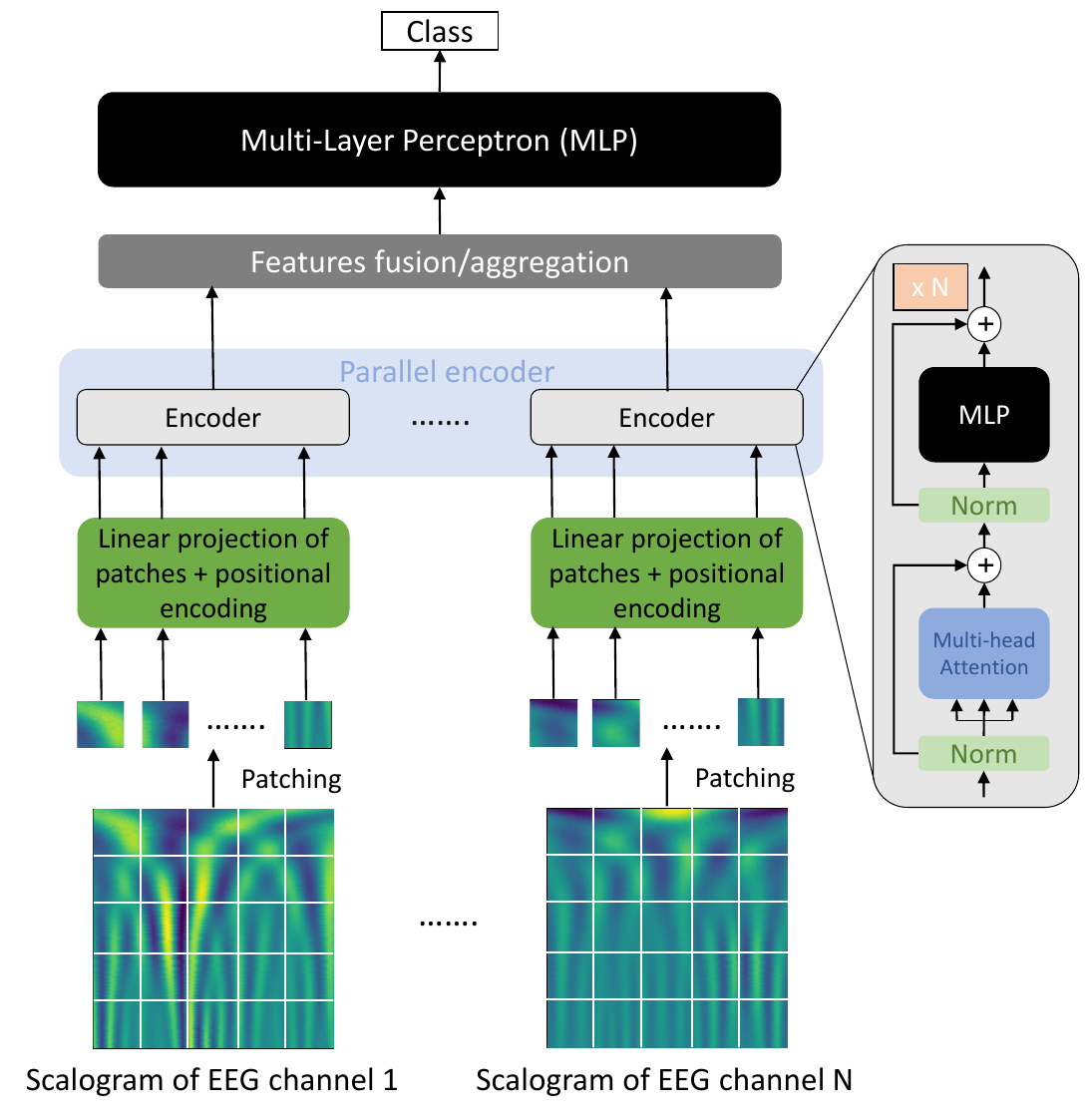}
    \caption{Schematic representation of the MViT architecture for EEG classification. The model ingests scalograms of individual EEG channels, processes them through parallel encoders, and fuses their features for final classification via a MLP. Scalograms are generated using the CWT to provide a time-frequency representation of the EEG data.}
    \label{FIG2}
\end{figure}

The MViT operates in a similar manner, except that the single encoder is replaced by an array of independent encoders, each dedicated to process one of the EEG channel. After these encoders have performed their individual tasks, their outputs are concatenated together before passing through the MLP classifier. As this model only works with two-dimensional data, the 1D EEG signal must be converted. This transformation is achieved in \cite{hussein2022} using the Continuous Wavelet Transform (CWT), a mathematical technique that decomposes the signal into its various frequency components at different time scales \cite{mallat1999}. This results in a time-frequency (2D) representation of the EEG data. Fig. \ref{FIG2}  provides a visual representation of the MViT architecture, where EEG scalograms (\textit{i.e.}, the output of the CWT) serve as input, and the model predicts their respective classes.

\subsection{Model specifications}
\subsubsection{Scales} To adapt to the size and purpose of each dataset, we designed two models of different scales from the architecture previously described. Their parametrizations are summarized in Table \ref{TAB04}. The EO/EC model was designed to be considerably smaller than its TUSZ counterpart for two reasons. First, reducing the model size allows to train more copies of it in a limited amount of time, resulting in the possibility to achieve significance testing on its performances. The task of eyes open and eyes closed classification is also rather simple and performed on a limited dataset, such that training a more consequent model would induce more overfitting.

\subsubsection{Optimizer and regularization} Both models underwent training using an AdamW optimizer \cite{losch2017} set with parameters $\beta_1 = 0.9$, $\beta_2 = 0.999$, weight decay $\lambda_{weights} = 10^{-4}$, and fixed learning rate $l_r = 10^{-4}$. Further regularization was applied in the form of dropout with a rate of 0.5 in the decision head and 0.1 inside the encoders.

\begin{table}[!t]
    \centering
    \caption{Hyper-parameters related to the model architecture and the number of weights this configuration yields.}
    \resizebox{.49\textwidth}{!}{\begin{tabular}{m{1.75 in}cc}
        \toprule
        \toprule
        Hyper-parameter & EO/EC & TUSZ \\
        \midrule
        Input tensor shape & [32, 25, 8] & [20, 25, 40] \\
        \# encoders in parallel & 32 & 20\\
        \# transformer layers in encoder & 1 & 8\\
        \# attention heads per layer & 2 & 4\\
        Size of hidden and output layers \newline (Transformer MLP) & [16, 8] & [80, 40]\\
        Size of hidden layers (Decision head) & [128, 64] & [512, 256]\\
        \midrule
        Weights & EO/EC & TUSZ \\
        \midrule
        \# trainable weights in parallel encoder & 28 160 & 5 248 000\\
        \# trainable weights in decision head & 827 665 & 10 372 177\\
        \# total trainable weights & 855 825 & 15 620 177\\
        \bottomrule\bottomrule
    \end{tabular}}
    \label{TAB04}
\end{table}

\section{Performances assessment}
\subsection{Benchmarking the pre-trainings} \label{sec:bench}
To identify the most effective data alteration method among the three discussed in Section \ref{SEC:Alt}, we utilized each of them to pre-train our MViT model on the EO/EC dataset. Leveraging the dataset's small size, we conducted our experiment 17 times, starting with a different seed for each occurrence. This repetition allowed us to establish the significance of the results through statistical testings.

\begin{figure}[!t]
    \centering
    \includegraphics[width=3.5in]{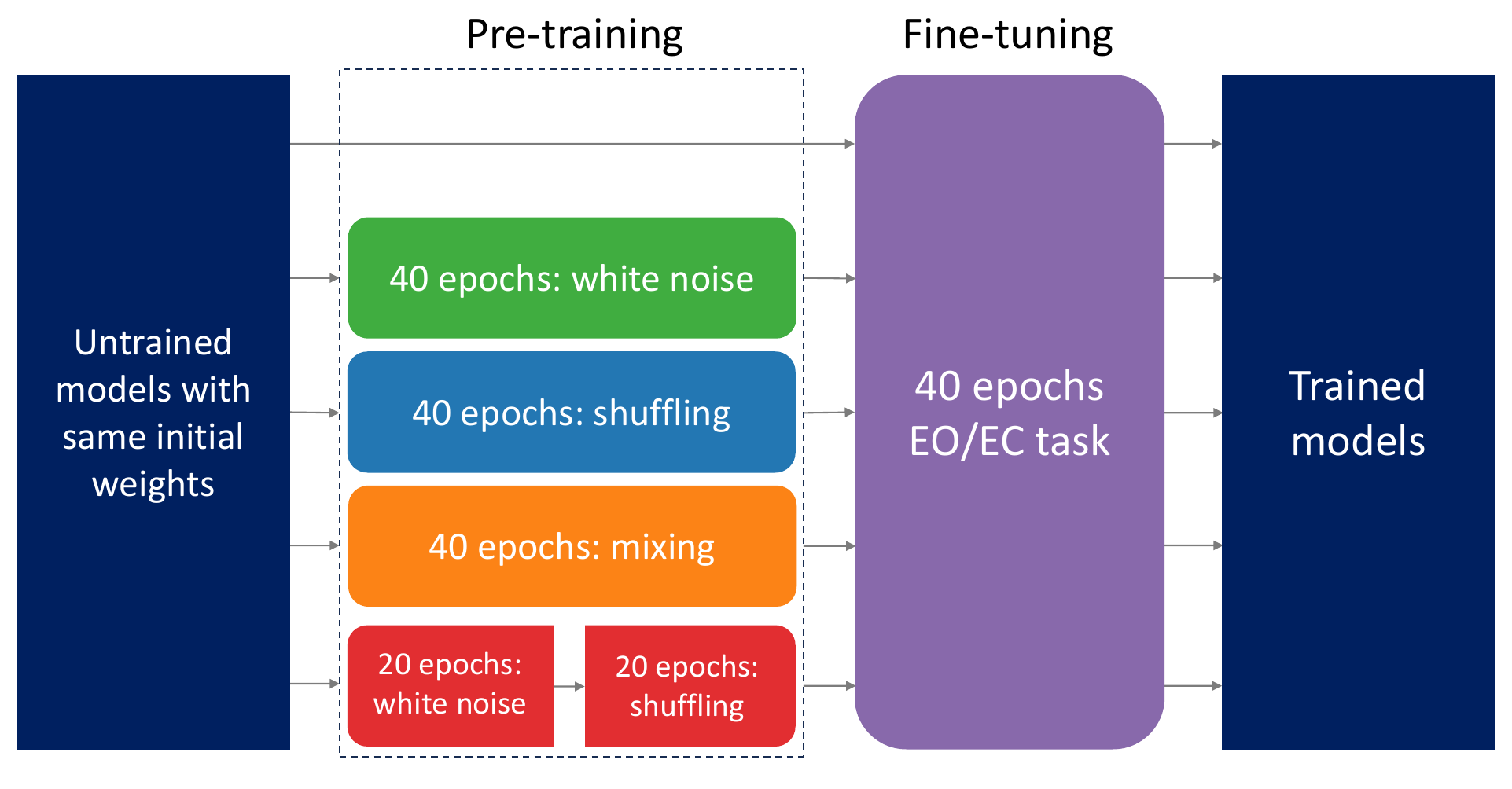}
    \caption{Pre-training performances assessment methodology on the designed pre-training datasets. Starting with five copies of a same untrained model, four of them are pre-trained separately with one of the proposed method for 40 epochs, while the fifth model is not pre-trained. All models are then fine-tuned on the specific task. This operation is repeated $N$ times (in this paper, $N=17$) to account for the variability between the experiments.}
    \label{FIG3}
\end{figure}

In each experiment, we employed three versions of the same model, all initialized with identical random weights. Each model was pre-trained on one of the three datasets created with the data alterations for 40 epochs. Subsequently, fine-tuning for the EO/EC classification task continued for an additional 40 epochs. We introduced a control model that underwent no pre-training, only fine-tuning for 40 epochs, commencing with the same initial weights as the other models. Additionally, we examined a fifth model with hybrid pre-training, consisting of 20 epochs of training on the dataset with white noise alteration, followed by 20 epochs on the dataset with channel shuffling alteration. We hypothesized that this hybrid pre-training would impart a broader range of features to the model, as white noise replacement focuses on teaching frequency content, while channel shuffling emphasizes channel correlations. A schematic representation of the pre-training performance assessment methodology is presented in Fig. \ref{FIG3}.

Four performance metrics were extracted at the end of each experiment, following model fine-tuning. These metrics include 1) the epoch of convergence (EOC), which denotes the epoch with the lowest validation loss; 2) the minimum validation loss; 3) the validation accuracy at the EOC; and 4) the validation area under the curve (AUC) at the EOC.

\subsection{Performances comparison in practical scenario} \label{sec:bench2}
Following the determination of the best pre-training method detailed in the previous section, our next step is to assess its effectiveness in a practical scenario, specifically, forecasting epileptic seizures. The objective here is to train our algorithm to distinguish between pre-ictal EEG patterns (revealing an imminent seizure) and inter-ictal EEG patterns (\textit{i.e.}, no imminent seizure) recorded from epileptic patients.

To assess the performance gains attributed to our proposed pre-training, we undertake a comparison between two identical models, each initialized with different weight configurations. The first model, referred to as "NPT" for "Non-Pre-Trained," begins with random weights initialization. The second model, termed "PT" for "Pre-Trained," undergoes pre-training on the most performant of our designed datasets. The weights from the model's epoch of convergence (EOC) during pre-training are then adopted as the initial weights for the primary task.

Both models are fine-tuned using the same training and validation data splits from the TUSZ dataset, and we employ an early stopping mechanism with a patience of 5 epochs to ensure that the loss of both models converges. As in section \ref{sec:bench}, we evaluate both models using four performance metrics: validation loss, validation accuracy, and validation AUC, all measured at the EOC, plus the EOC itself. Additionally, we measure the loss, accuracy, and AUC of both models on the test split, resulting in a total of 7 performance metrics.

\section{Results and discussion}
\subsection{Benchmarking the pre-trainings}
\begin{figure}[!t]
    \centering
    \includegraphics[width=3.5in]{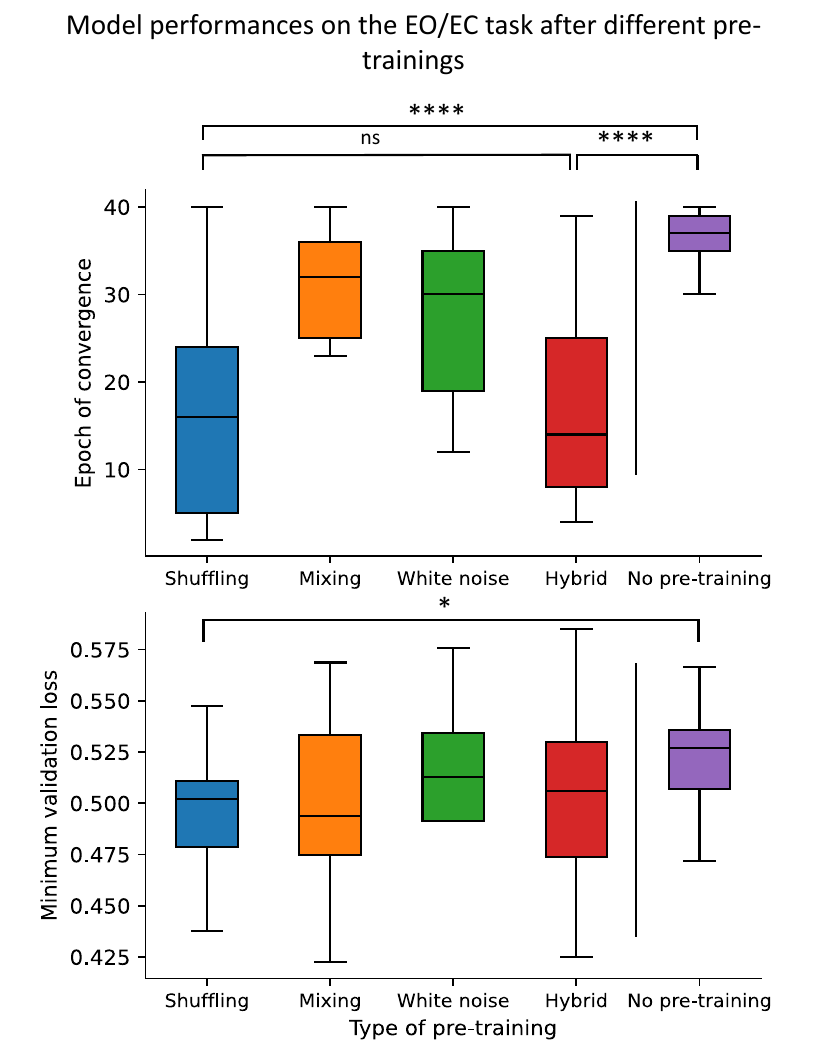}
    \caption{Box plots of the EOC (top) and the minimum validation loss  (bottom) for each of the proposed pre-training methods. The significance levels are: $p< 0.05$ (*), $p<0.01$ (**), $p<10^{-3}$ (***), and $p<10^{-4}$ (****) (not all relations of significance are shown).}
    \label{FIG4}
\end{figure}

Fig. \ref{FIG4} displays a box plot showing the distribution of the EOC and the minimum validation loss across all pre-training methods. Additionally, Table \ref{TAB01} provides a summary of the previously selected performance metrics, including means and standard deviations, for each pre-training approach.

To assess the statistical significance of eventual differences in performances, we conducted two-tailed Welch's t-tests \cite{welch1947} on each pair of pre-training methods. This test was selected based on the assumption that the experimental results can be considered independent and identically distributed (iid) random variables, and that the true variances of the samples differ. The summary of the significance testing is added to Table \ref{TAB01}.

\begin{table}[!t]
    \centering
    \caption{Mean performances for each type of pre-training on the EO/EC dataset. Values in parentheses are the standard deviations. The significance of the performance improvement with regards to the non-pre-trained model was tested at $p< 0.05$ (*), $p<0.01$ (**), $p<10^{-3}$ (***), and $p<10^{-4}$ (****).}
    \resizebox{.49\textwidth}{!}{\begin{tabular}[width=3.5in]{lcccc}
        \toprule\toprule
        Pre-training & EOC &  Min val. loss & Val. acc. [\%] & Val. AUC \\
        \midrule
        White noise & 27.4 (9.58)** & 0.497 (0.053) & 76.8 (4.92) & 0.847 (0.036)\\
        Shuffling & 17.0 (12.0)**** & 0.500 (0.029)* & 77.4 (2.74) & 0.839 (0.026)\\
        Mixing & 31.4 (5.37)** & 0.496 (0.040) & 77.8 (4.78) & 0.843 (0.035)\\
        Hybrid & 17.2 (10.9)**** & 0.507 (0.050) & 78.3 (4.35) & 0.828 (0.032)\\
        \midrule
        Pooled & 23.2 (11.6)**** & 0.500 (0.044)* & 77.6 (4.32) & 0.839 (0.033)\\
        \midrule
        No pre-training & 36.1 (4.27) & 0.520 (0.025) & 75.7 (4.36) & 0.825 (0.025)\\
        \bottomrule\bottomrule
    \end{tabular}}
    \label{TAB01}
\end{table}

When comparing the pooled pre-training performances with the ones of the model without pre-training,  we observed no significant improvements in terms of validation accuracy and validation AUC at the EOC, as indicated by p-values exceeding 0.05 (p-values: 0.1419 and 0.0661, respectively). However, we observed highly significant improvements in EOC (p-value $<10^{-5}$) and significant improvements in minimum validation loss (p-value = 0.0197) with pre-training.

Among the pre-training methods, only the shuffling alteration demonstrated a significant improvement in minimum validation loss compared to the non-pre-trained model (p-value = 0.0441). While the statistical tests indicated a trend for other alteration types (p-values: 0.0530 for channel mixing and 0.1329 for white noise replacement), these trends did not cross the significance threshold $\alpha = 0.05$. The hybrid pre-training approach showed the least significant difference in this metric, with a p-value of 0.3474.

When examining the EOC, both the shuffling alteration and the hybrid pre-training approach exhibited highly significant improvements compared to the non-pre-trained model (p-value $<10^{-5}$ for both). Moreover, they were shown to perform significantly better than channel mixing (p-values: 0.0024 for shuffling and 0.0011 for hybrid) and white noise replacement (p-values: 0.0132 for shuffling and 0.0088 for hybrid). However, it is worth noting that the standard deviation of the EOC for these two pre-training methods is higher than that of the other alterations.

To investigate a potential trade-off between the speed of convergence (determined by the EOC) during pre-training and the minimum validation loss, we conducted a linear regression analysis. The results (Table \ref{TAB02}) revealed that the EOC explained only a small portion of the variance in the minimum validation loss, with the highest $R^2$-score being slightly above 0.25. This led to the conclusion that earlier convergence during pre-training did not compromise the validation loss.

\begin{table}[!t]
    \centering
    \caption{Slopes and $R^2$ scores of the linear regressions fitting the EOC-minimum validation loss relation. The pre-trainings are ranked from the lowest to the highest $R^2$.}
    \resizebox{.4\textwidth}{!}{\begin{tabular}{lcc}
        \toprule\toprule
        Pre-training & Slope [loss increase/epoch]  & $R^2$ \\
        \midrule
        Mixing & $1.049\times 10^{-3}$ & 0.0194\\
        Shuffling & $-0.488 \times 10^{-3}$ & 0.0399\\
        No pre-training & $-1.637\times 10^{-3}$ & 0.0763\\
        White noise & $-2.459\times 10^{-3}$ & 0.1985\\
        Hybrid & $-2.351\times 10^{-3}$ & 0.2581\\
        \bottomrule\bottomrule
    \end{tabular}}
    \label{TAB02}
\end{table}

Based on the gathered results, datasets with the shuffling alteration proved to be the most effective for pre-training, significantly improving both EOC and minimum validation loss. Therefore, a dataset with this alteration pre-trained the model in the seizure forecasting task.

The hybrid pre-training approach, although significantly reducing the EOC, did not exhibit a significant improvement in the minimum validation loss. Exploring alternative methods of hybrid pre-training, such as applying multiple alterations simultaneously or mixing multiple datasets, could therefore be an avenue for future research.

\subsection{Performances comparison in practical scenario}
Building upon the insights gained in the previous section and following the protocol established in section \ref{sec:bench2}, we initiated the pre-training of our larger transformer model using a dataset generated by means of the shuffling alteration, applied to a portion of the TUSZ data.

When comparing the time necessary for the Pre-Trained (PT) and Non-Pre-Trained (NPT) models to converge to a local validation loss minimum (Table \ref{TAB03}), it clearly appears that pre-training allows for a much faster fine tuning, as the number of epochs required to find a minimum is halved (going down from $8$ epochs to $4$) when the only difference between the models is whether the starting weights are random or issued from pre-training. To put it differently, once the initial cost of pre-training the model on an important amount of data is incurred, it will serve as a foundational point for various related tasks, thereby reducing their fine-tuning duration.

\begin{table}[!t]
    \centering
    \caption{Compilation time of both models for the different training phases.}
    \resizebox{.49\textwidth}{!}{\begin{tabular}{lccc}
        \toprule\toprule
        Training phase & Time/epoch (avg.) [s] & EOC & Tot. train. time [h]\\
        \midrule
        Pre-training (PT) & 11 117 & 9 & 27.8\\
        Fine-tuning (PT) & 5 288 & 4 & 5.88\\
        Fine-tuning (NPT) & 5 698 & 8 & 12.66\\
        \bottomrule\bottomrule
    \end{tabular}}
    \label{TAB03}
\end{table}

Turning our attention towards model performance, the evaluation of PT and NPT, summarized in Table \ref{TAB10}, reveals improvements in all assessed metrics when employing our pre-training method before fine-tuning. In particular, the test loss reduces from $0.1939$ in the absence of pre-training to $0.1748$ when initializing with pre-trained weights. This reduction has a positive cascading effect on test accuracy and AUC.

Taken together, these findings indicate that the pre-training datasets developed in this study not only accelerate model convergence during task-specific training but also contribute to the creation of models with better classification performance compared to models lacking pre-training.

\begin{table}[!t]
    \centering
    \caption{Comparison of the model performances after fine-tuning depending on whether it was pre-trained or not.}
    \resizebox{.4\textwidth}{!}{\begin{tabular}{lcc}
        \toprule\toprule
        Metric & PT &NPT\\
        \midrule
        Validation loss at EOC & 0.1689 & 0.2035\\
        Validation accuracy at EOC [\%] & 92.15 & 90.43\\
        Validation AUC at EOC & 0.9719 & 0.9630\\
        \midrule
        Test loss & 0.1748 & 0.1939\\
        Test accuracy [\%] & 92.16 & 90.93\\
        Test AUC & 0.9702 & 0.9648\\
        \bottomrule\bottomrule
    \end{tabular}}
    \label{TAB10}
\end{table}

\section{Conclusion and future works}
In this study, we devised three distinct methods for pre-training deep learning models on EEG tasks using unlabeled data. These techniques all revolve around the core task of EEG identification within a variety of multi-channel signals, aiming to impart the model with essential intrinsic properties of EEG, such as its characteristic frequency behavior and channel correlations.

Our findings demonstrate a consistent and significant improvement in training time for models pre-trained with any of the designed pre-training methods compared to models initialized with random weights. Additionally, one specific pre-training approach, involving the shuffling of EEG channels and training the model to distinguish shuffled from unaltered EEG data, stands out by substantially enhancing minimum validation loss when compared to non-pre-trained models.

The applicability of this specific pre-training method was further validated in the context of a seizure forecasting task, leveraging the Temple University Seizure Detection Corpus. In this task, the pre-trained model not only exhibited significantly faster training, converging to optimal performance in less than half the epochs required by an identical non-pre-trained model, but also achieved superior results in terms of loss, accuracy, and AUC for both the validation and test sets.

Moving forward, future researches building upon the results of this study should explore the performances of the proposed pre-training techniques on a broader range of EEG datasets and applications. In particular, there's a need to assess the effectiveness of these pre-training methods for different models and architectures, with a focus on the less data-greedy ones.

In conclusion, our research underscores the significance of pre-training in the context of EEG signal analysis, while also exploiting more easily accessible and otherwise useless unlabeled data. These findings open doors to the exploitation of more data-greedy architectures in the field of EEG classification and allows to build general models whose weights can serve as a basis for many specific task requiring EEG analysis.

\section*{Code availability}
The code developed to generate the pre-training datasets from unlabeled data is available at the following GitHub repository: \href{https://github.com/tbary/EEGPreTrainingDatasets}{https://github.com/tbary/EEGPreTrainingDatasets}. Researchers can access and utilize this code for further investigation and experimentation.

\bibliographystyle{IEEEtran}
%
\newpage
\bibliography{biblio.bib}

\end{document}